# Development and preliminary tests of resistive microdot and microstrip detectors

**P. Fonte,**[a] **E. Nappi**[b]**, P. Martinengo**[c]**, R. Oliveira**[c]**, V. Peskov**[c,d,*]**, F. Pietropaolo**[e]**, P. Picchi**[f]

[a] *LIP and ISEC, Coimbra, Portugal*
[b] *INFN Bari, Bari, Italy*
[C] *CERN, Geneva-23, Switzerland, CH -1211*
[d] *UNAM, Mexico City, Mexico*
[e] *INFN Padova, Padova, Italy*
[f] *INFM Frascati, Frascati, Italy*
*E-mail*: Vladimir.peskov@cern.ch

ABSTRACT: In the last few years our group have focused on developing various designs of spark-protected micropattern gaseous detectors featuring resistive electrodes instead of the traditional metallic ones: resistive microstrip counters, resistive GEM, resistive MICROMEGAS. These detectors combine in one design the best features of RPCs (spark-protection) and micropattern detectors (a high position resolution).
In this paper we report the progress so far made in developing other types of resistive micropattern detectors: a microdot-microhole detector and a microgap-microstrip detector. The former detector is an optimal electron amplifier for some special designs of dual phase noble liquid TPCs, for example with a CsI photocathode immersed inside the noble liquid. Preliminary tests of such a detector, for the first time built and investigated, are reported in this paper. The latter detector is mainly orientated towards medical imaging applications such as X-ray scanners. However, we believe that after a proper gas optimization, these detectors could also achieve a high time resolution and could thus be used in applications as TOF-PET, detection of charged particles with simultaneous high time and position resolution etc.

KEYWORDS: micropattern detectors; microgap RPC, CsI, dual phase noble liquid detectors, X-ray scanners.

* Corresponding author.

## 1. Introduction

Micropattern gaseous detectors (microstrip detectors, MICROMEGAS, GEMs and others) offer an unprecedented 2D position resolution (20-40 μm) making them very attractive for many applications such as: tracking of charged particles, detection and visualization of X-rays and UV photons. However, owing to the fine structure of their electrodes, micropattern sensors can be easily damaged by sparks, which are almost unavoidable under real experimental conditions. Sparks appear when the total charge in the avalanche approaches the critical value $Q_{crit}$, typically $10^6$-$10^7$ electrons, depending on the specific design of the detector [1]. In some cases, however, at total charge values in the avalanche $Q_s<Q_{crit}$, surface streamers may form along the dielectric surfaces between the anode and the cathode electrodes and subsequently might also trigger harmful discharges [2]. Usually surface streamers appear if the field lines are parallel to the dielectric surface (along all the surfaces) between the anode and the cathode electrodes.

At the last three RPC conferences [3-5], we reported that the resistive electrode approach, used in RPCs, can be successfully applied to micropattern gaseous detectors making them spark-protected. Our first spark-protected detectors - GEM and MICROMEGAS - had unsegmented resistive electrodes [3, 5]. Lately we have focused on the development of sensors with segmented (strips) resistive electrodes (see for example [6]).This approach turned out to be very fruitful allowing the development of various designs of large-area spark-protected micropattern detectors tailored to various applications (see [7] and references therein). For instance, recently, MICROMEGAS with resistive anode strips were developed for the ATLAS forward wheel upgrade (R-MICROMEGAS) [8].

In this paper, we will introduce two new designs of resistive microstrip detectors: the Microdot–Microhole Detector (RMMD) and the Microgap-Microstrips-RPC (MMRPC). These innovative micropattern spark-protected detectors are primarily orientated towards applications in which some members of our team are deeply involved: a dual phase noble liquid Time projection Chambers (TPCs) with a CsI photocathode immersed inside the noble liquid and a TOF-PET, and X-ray imaging devices [7, 9, 10]. Consequently the paper consists of two main parts: one dedicated to the RMMD for dual phase noble liquid TPCs and the other one dealing with MMRPCs for medical applications.

## 2. Microdot–Microhole Detector for dual phase noble liquid TPCs

### 2.1 A prototype of a dual-phase LAr detector with a CsI photocathode

Our group is involved in developing the innovative concept of a dual phase noble liquid TPC for dark matter search. The principle of operation of such a detector is described in many papers (see, for example, a recent review [11]). It is basically a noble liquid TPC housing two phases (liquid-gas) of a noble element in a single cell. The operational principle is as follows: any interaction inside the volume of the noble liquid will create an ionization track there (with $n_0$ primary electrons). Some fraction of the created ions $n_0\eta_r$ will recombine and create a flash of scintillation light which is detected by photomultiplier tubes (PMTs) surrounding this volume. Because this detector is operating as a TPC, with an applied drift electric field, some fraction of ions and electrons $n_0\eta_d$ will escape recombination and free electrons will drift along the field lines towards the border between the liquid and gaseous phases, which they reach after a time $t_d$. As was first shown in the work of Dolgoshein et al [12], in a rather strong electric field applied across the border (~10kV/cm), these electrons can be extracted into the gas phase and subsequently detected by using two parallel-meshes where the electrons produce secondary scintillation light, the intensity of which is proportional to the amount of extracted charges. The ratio of the first to the second scintillation light depends on the nature of the interaction. For example, in the case of the detection of recoil tracks with high density of ionization $\eta_r >> \eta_d$; whereas, in the case of the detection of gamma radiation with low density tracks, $\eta_r \leq \eta_d$. Hence, by measuring the ratio of the scintillation lights one can discriminate between various interactions and select only desirable events.

As mentioned above, to efficiently detect the primary scintillation light the entire volume of the dual phase detector is usually surrounded by an array of PMTs, resulting into increases of the cost of such a device. For this reason several research groups are investigating the possibility of reducing the number of PMTs by using cost effective alternative detectors: GEM [13], TGEM [14], avalanche photodiodes, silicon PMT etc [11].

The goal of our studies is to test another possible approach: to replace some PMTs in noble liquid TPC by a CsI photocathode immersed inside the liquid. In the works of Aprile's team [15-17], it was discovered that a CsI photocathode immersed inside the noble liquids exhibits very high quantum efficiency comparable to that of the PMTs. To our best knowledge, however, this concept was not yet never realized in any experimental device. The first simplified prototype of a dual phase detector with a CsI photocathode inside, which was built and tested in our laboratory, is shown schematically in figure 1. It was housed in an ultra-high-vacuum stainless steel vessel, 65 cm in diameter and 170 cm in height, for a total volume of 550 liters. In order to control the heat losses, the whole vessel was partially embedded in a thermal bath of commercial LAr contained in an open air dewar. The cryostat was equipped with a standard recirculation/purification system, the filling of the main container being performed through an Oxisorb/Hydrosorb filter. This allowed a LAr recirculation rate of about 5 liters per hour with the purification absorber working in the gas phase. The initial LAr filling of the cryostat took about 2 h. An electron lifetime of several milliseconds was reached after a few days of filling.

Usually we condensed some amount of LAr (up to 100 liters) in the cryostat and maintained gaseous Ar above the liquid. The level of liquid was monitored with a capacitor-based sensor.

As in the case of any other standard dual-phase detector two parallel meshes were placed above the liquid volume. The gap between meshes was 1cm. A photomultiplier tube ETL9357 was placed above the parallel meshes, the window of the PMT was coated with a TPB layer for shifting the UV scintillation light into visible light in order to match the PMTs sensitivity range. Ionization inside the LAr was produced by an alpha source $^{241}$Am deposited on a tiny needle-shaped surface.

The main innovative feature of this detector was the employ of a CsI photocathode immersed inside the LAr consisting of a stainless steel disc of 10cm in diameter coated with a 0.4 μm thick CsI layer. A typical voltage applied between the meshes was 1.5kV, while the voltage applied to the CsI photocathode was -3kV.

**2.2 Preliminary measurements with the PMT**

Figure 2 shows the results of the measurements performed with the PMT tube when the cryostat was filled with Ar gas while figure 3 refers to the operations in dual-phase mode. In both cases alpha particles from the Am source produced primary UV light detected by the PMT (the first pulse in the oscillograms presented in Figures 2 and 3). Some of the created primary electrons move to the space between the grid and produce secondary light (second pulse on the oscillograms). As was described in the previous paragraph, in standard dual phase TPCs these two pulses provide important information about the ionization event. In our case, however, in contrast to the ordinary noble liquid TPC, several more pulses were observed (see figures 2 and 3). These additional pulses appear because both the primary and the secondary light reach the CsI photocathode, extract electrons from it, which drift to the space between the meshes and produce other bursts of secondary light. These feedback pulses are clearly seen when the cryostat is filled with gaseous Ar and they become even stronger in the case of LAr. Obviously, this is a drawback of this type of dual phase detector design. One of the ways to avoid feedback is to use a light /charge multiplication structure that is geometrically shielded from the CsI photocathode. In our previous paper [18] we proposed using a resistive microhole and strip plate (figure 4) as such a shielding structure. Later studies, however, indicated that higher gas gains can be achieved with a microdot detector [19]. For this reason, in this paper, we focus on this more attractive approach, but to implement it we have modified the designs by manufacturing microdot structures on the top of a thick resistive GEM plate.

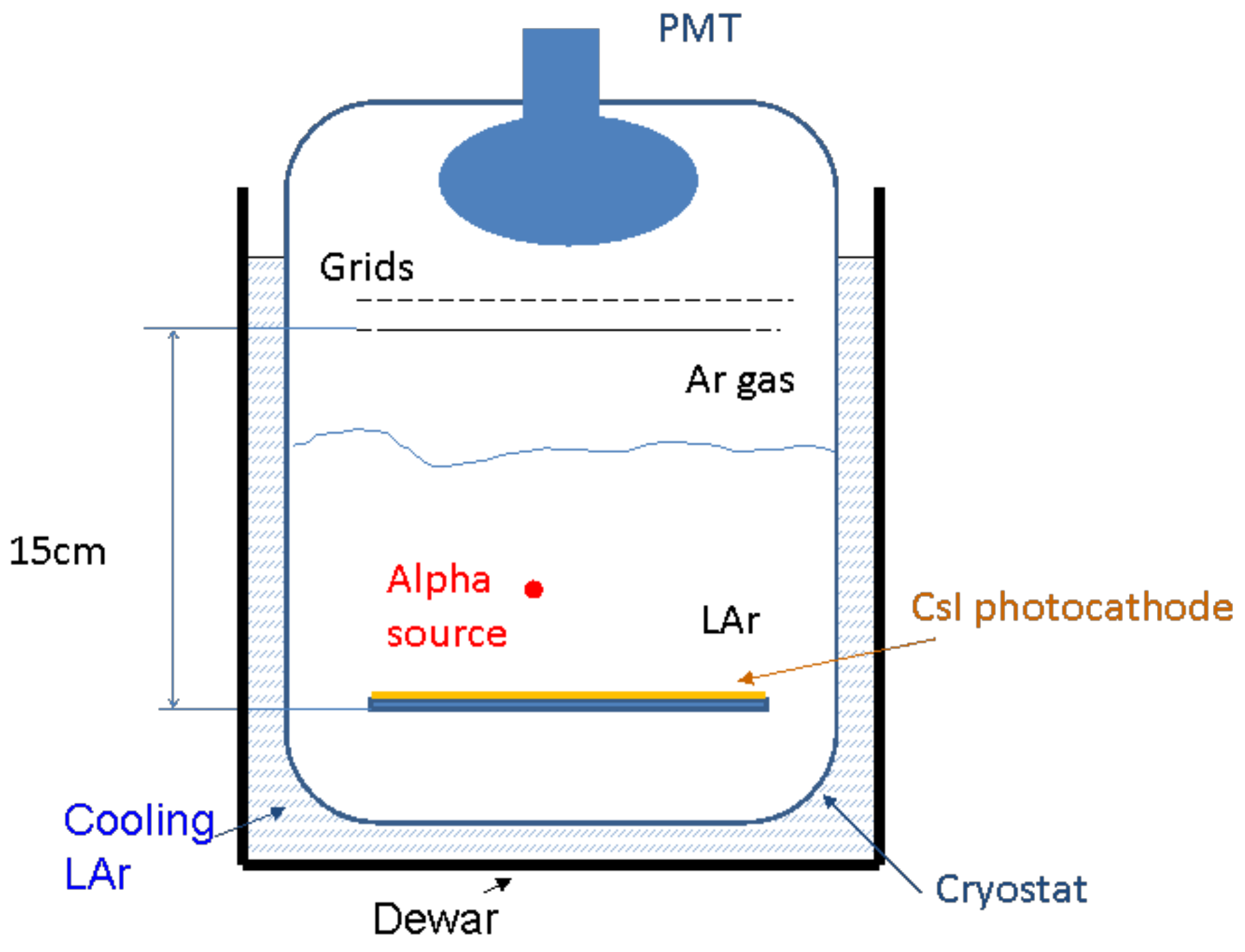

Figure 1. Schematic drawing of a dual phase LAr detector with a CsI photocathode immersed inside the liquid.

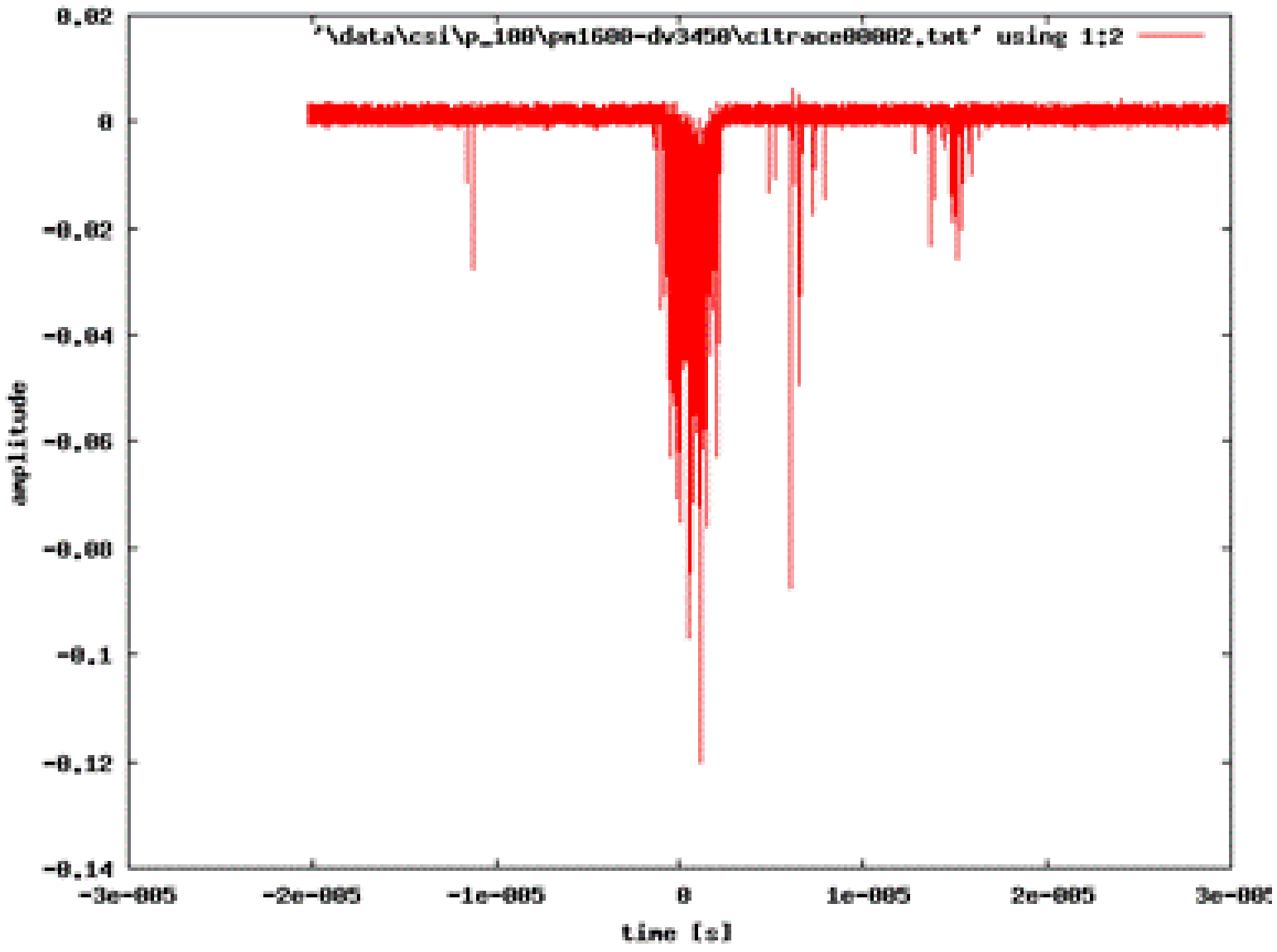

Figure 2. Oscillogram of signals from the PMT when the cryostat was filled with gaseous Ar. The first pulse is the primary scintillation light produced by alpha particles in Ar, the second peak is the secondary scintillation light produced by primary electrons between the meshes. The two other peaks are due to the photoelectrons extracted from the CsI photocathode.

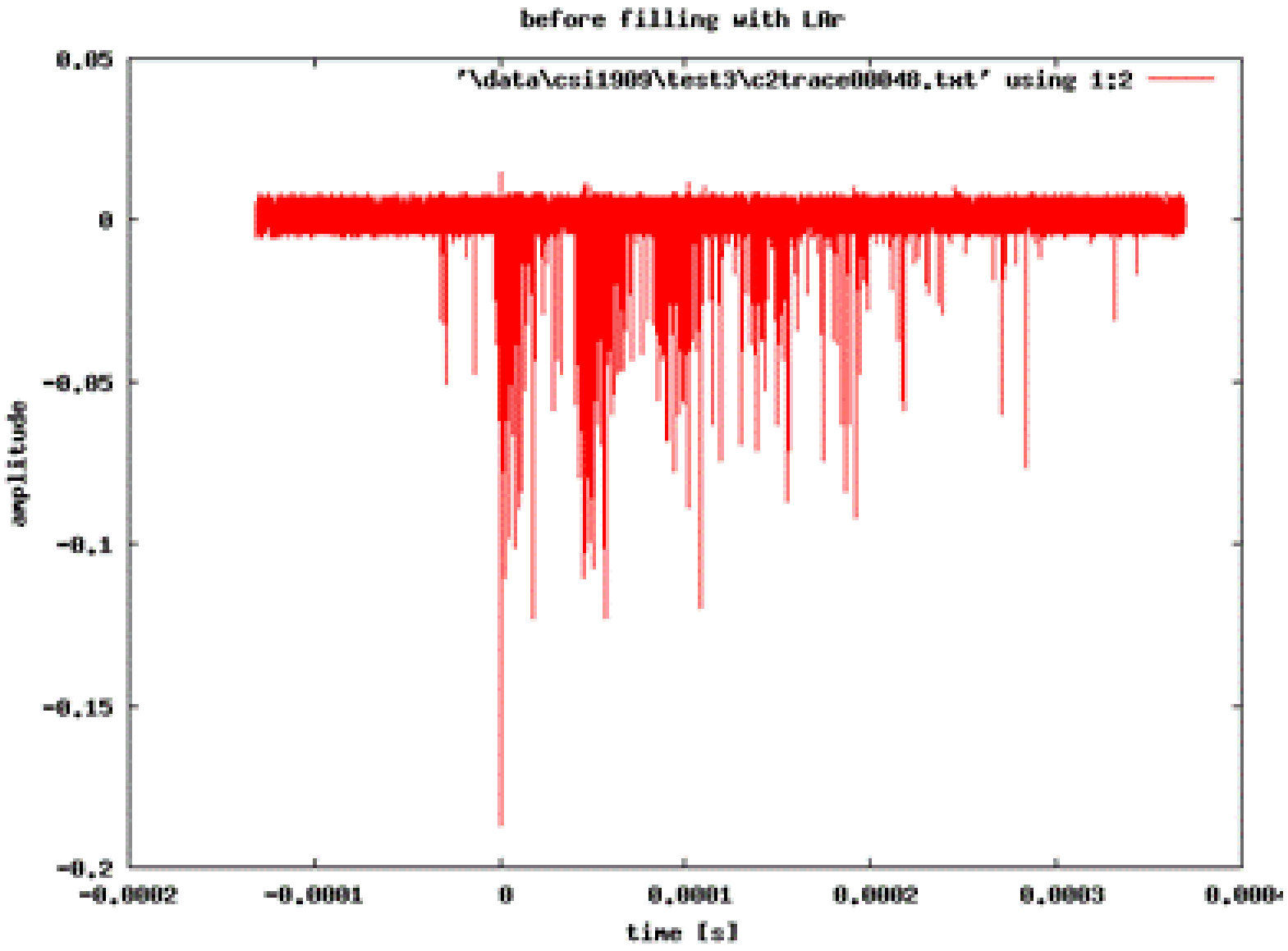

Figure 3. Oscillograms of the PMT signals delivered the dual phase LAr detector. As in the previous figure one can see the primary and secondary light peaks followed by feedback pulses.

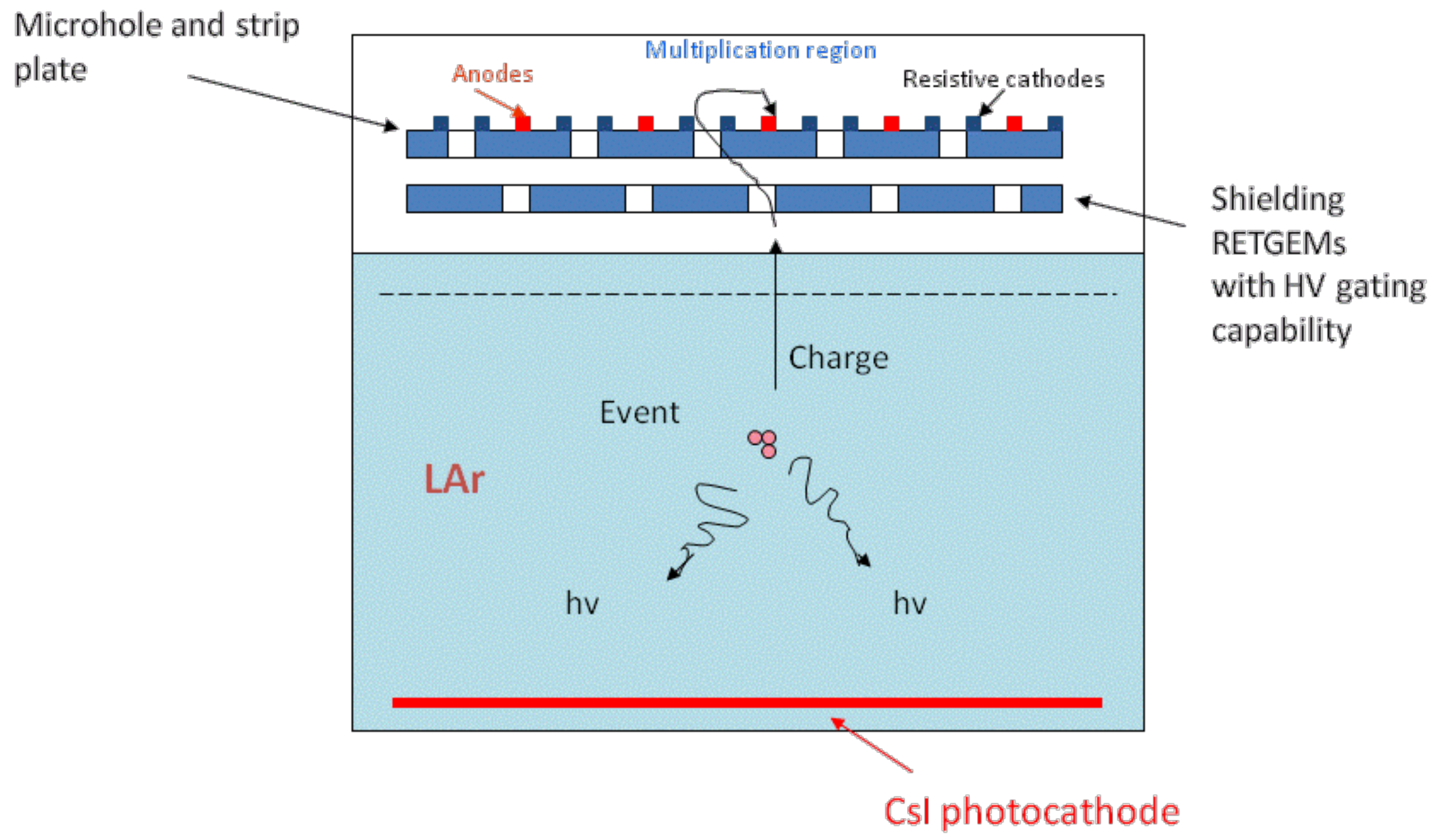

Figure 4. Design of a dual phase detector as proposed in [18], in which it was suggested to replace the parallel meshes with a resistive micro-hole microstrip detector. In this way, the light (or charge) multiplication region (strips) is geometrically shielded from the CsI photocathode.

### 2.3. RMMD for dual phase LAr detector with a CsI photocathode

#### 2.3.1 Manufacturing of RMMD

The RMMD was manufactured from a standard multilayer PCB board. The top layer of the PCB was covered by a 35 µm thick Cu layer, the inner layer of the PCB board contained Cu strips 100µm wide and with a pitch of 1mm (see figure 5a). On the top of the PCB board, parallel grooves, 100 µm wide, were created by photolithographic techniques (figure 5b). Subsequently, the grooves were filled with resistive paste ELECTRA polymers (figure 1c). Then the residual Cu strips on the top of the PCB were chemically etched (figure 1d) and the entire top surface was covered (laminated) with a DuPont Coverlay film. Part of the Coverlay layer was photochemically removed creating open dots. The Coverlay covers only the edges of the anode dots (diameter of 25 µm) and the edges of the cathode strips (Figure 1e). The thickness of the Coverlay layer on the top of resistive electrodes was 5 µm. Finally holes of 0.3 mm in diameter were drilled along the resistive cathode strips by a CNC machine (Figure 6, 7). A photograph of the RMMD is shown in figure 8.
The principle of operation of the RMMD is similar to the so-called microstrip-microhole detector with Cu electrodes [20. This geometry allows efficient suppression of photon and ion feedbacks [20]. Therefore, RMMDs are optimal for applications like photon detectors or TPCs.

RMMDs have two important advantages over the original microhole detectors [20]: 1) spark resistance, 2) the Coverlay structure prevents surface streamers from appearing thus allowing high gas gains to be achieved. As a result, the maximum achievable gas gain, $A_m$, is limited only by the value of $Q_{crit}$, $A_m=Q_{ctrit}/n_0$, where $n_0$ is being the number of primary electrons created in the detector's drift region by the radiation. High gas gains are essential in the case of the noble liquid TPC, the aim of which is to detect several primary electrons.

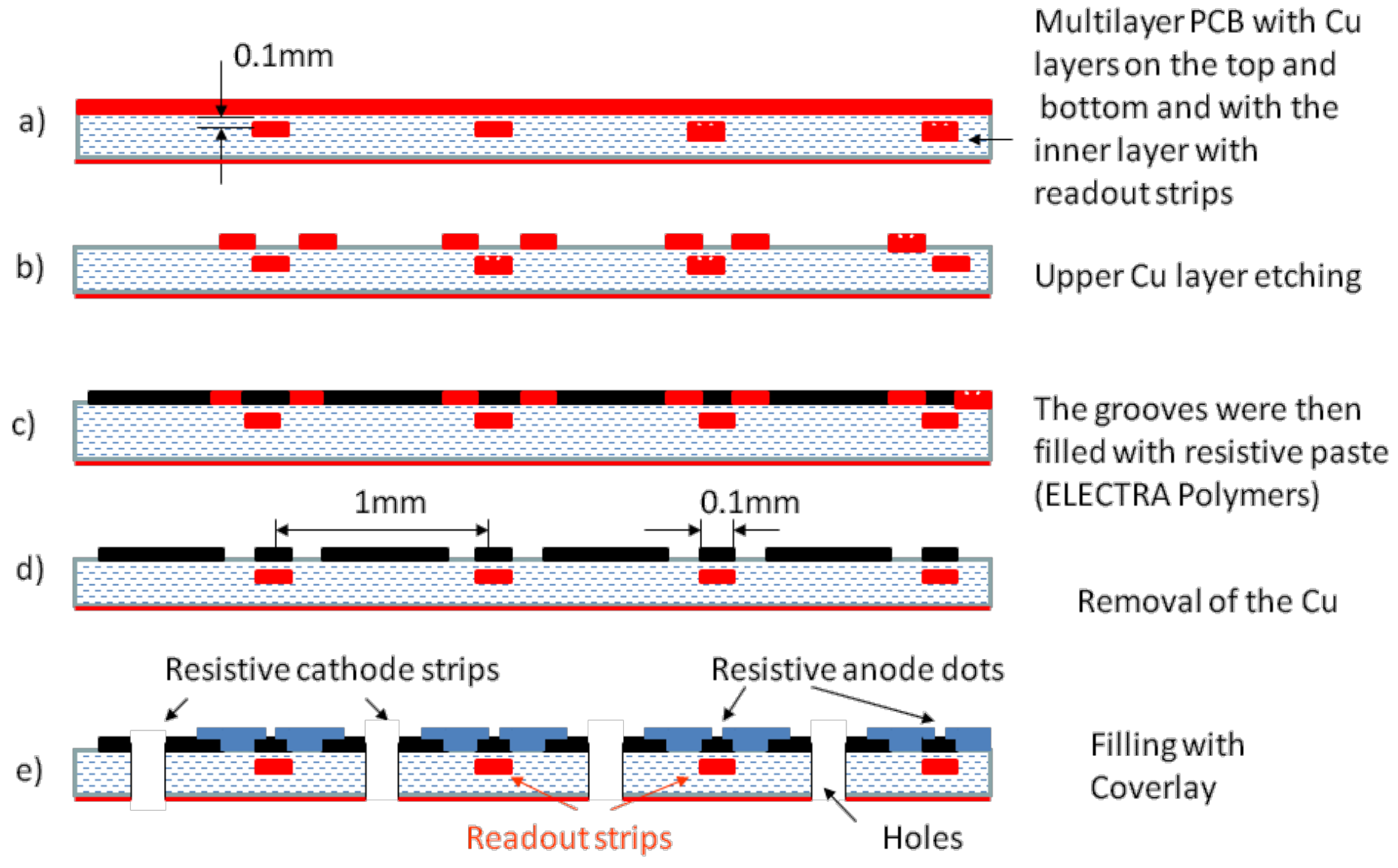

Figure.5. Manufacturing procedure of a RMMD.

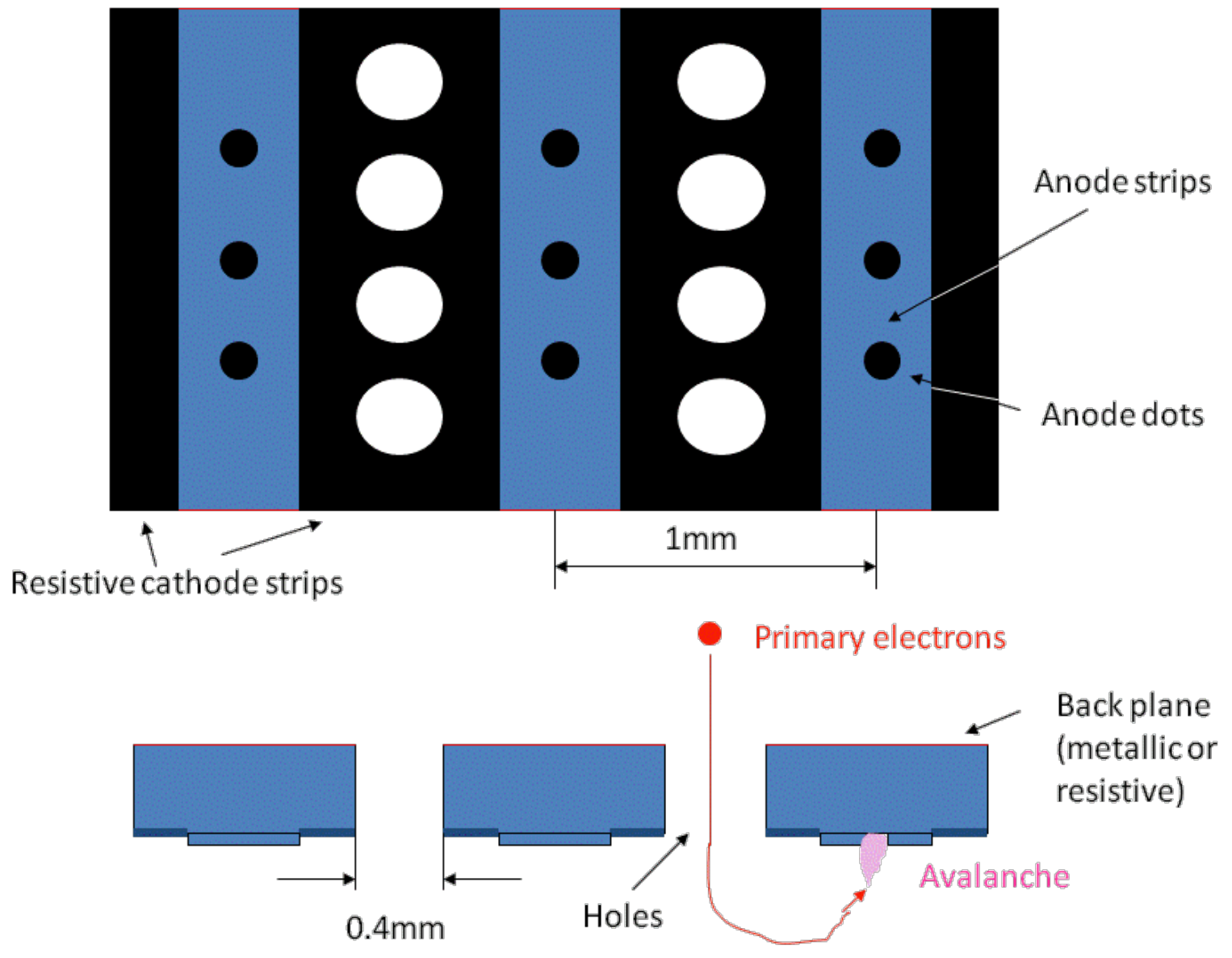

Figure 6. Technical drawing of the RMMD.

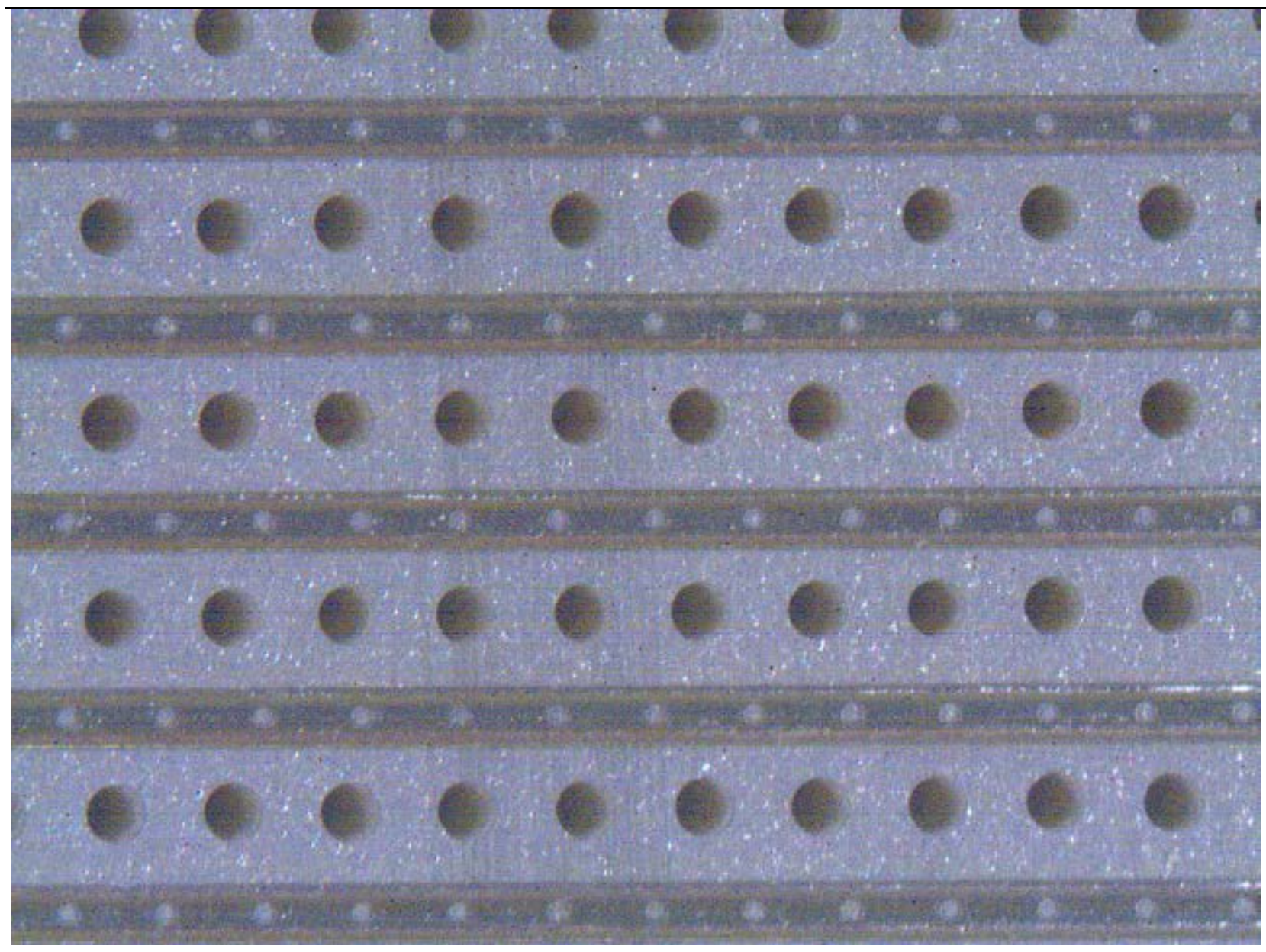

Figure 7. Magnified photograph of the RMMD manufactured on the top of resistive GEM. One can see GEM holes and rows of anode dots produced between the holes.

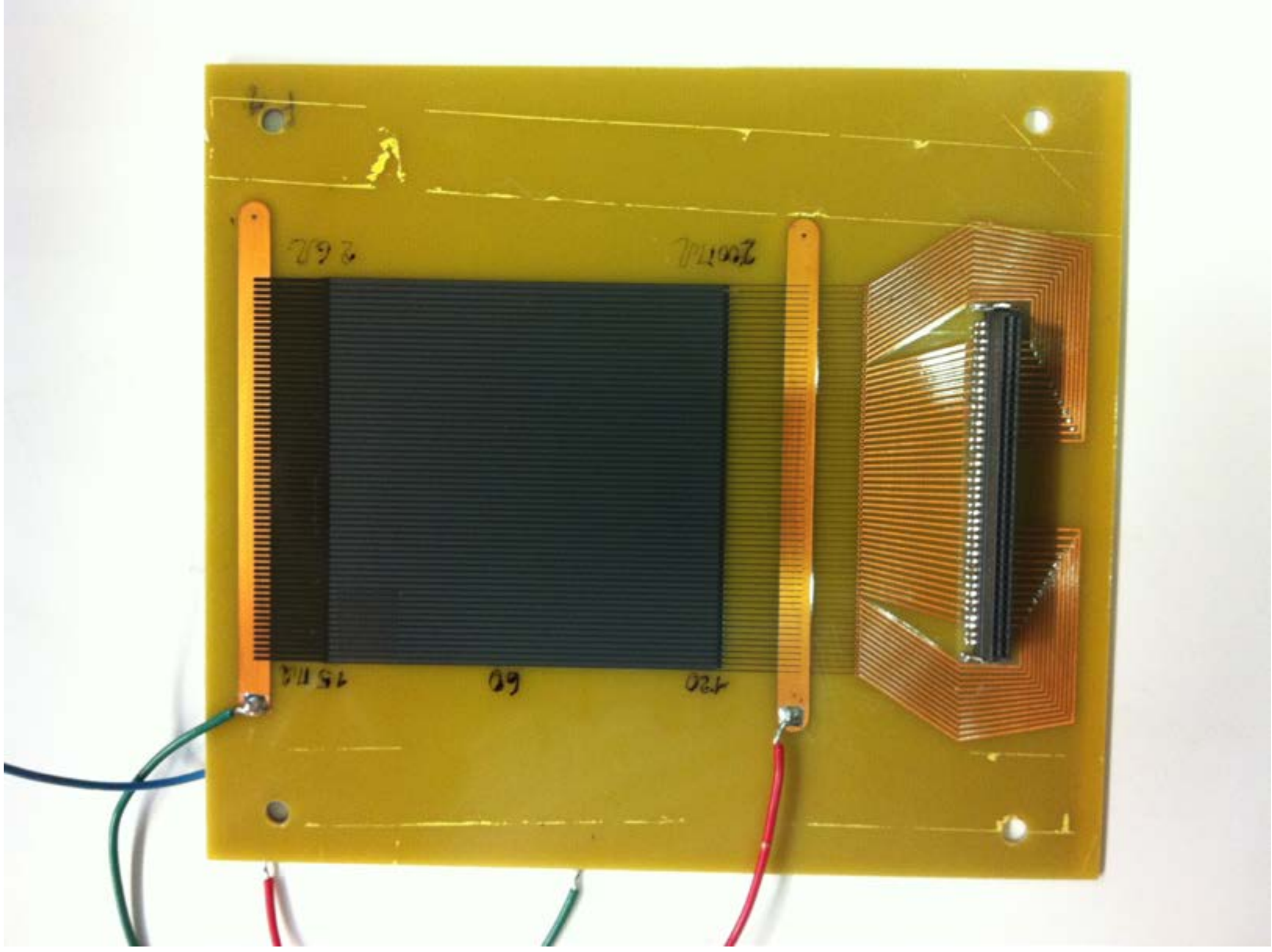

Figure 8. Photograph of one of the prototypes of the RMMD.

### 2.3.2. A simplified experimental setup set up for tests of RMMD

The cryostat described in the previous section was too bulky (recall that its volume was 500 liters and its weight over 100 kg) for the preliminary tests of RMMD. For this reason we devised a much simpler setup shown in figure 9. Most of the tests were done in Ar at a total pressure of 1atm. The primary electrons in the detector volume were created using one of the following sources: $^{241}$Am (alpha particles), $^{55}$Fe (5.9 keV photons) or by an Hg UV lamp. If necessary, the gas chamber with the detector inside could be cooled to cryogenic temperatures to perform gas gain measurements under these conditions. The signals from the anode dots of the RMMD were recorded by a charge sensitive amplifier Ortec 142pc. In some tests we also performed current measurements using a picoammeter Kethley 487.

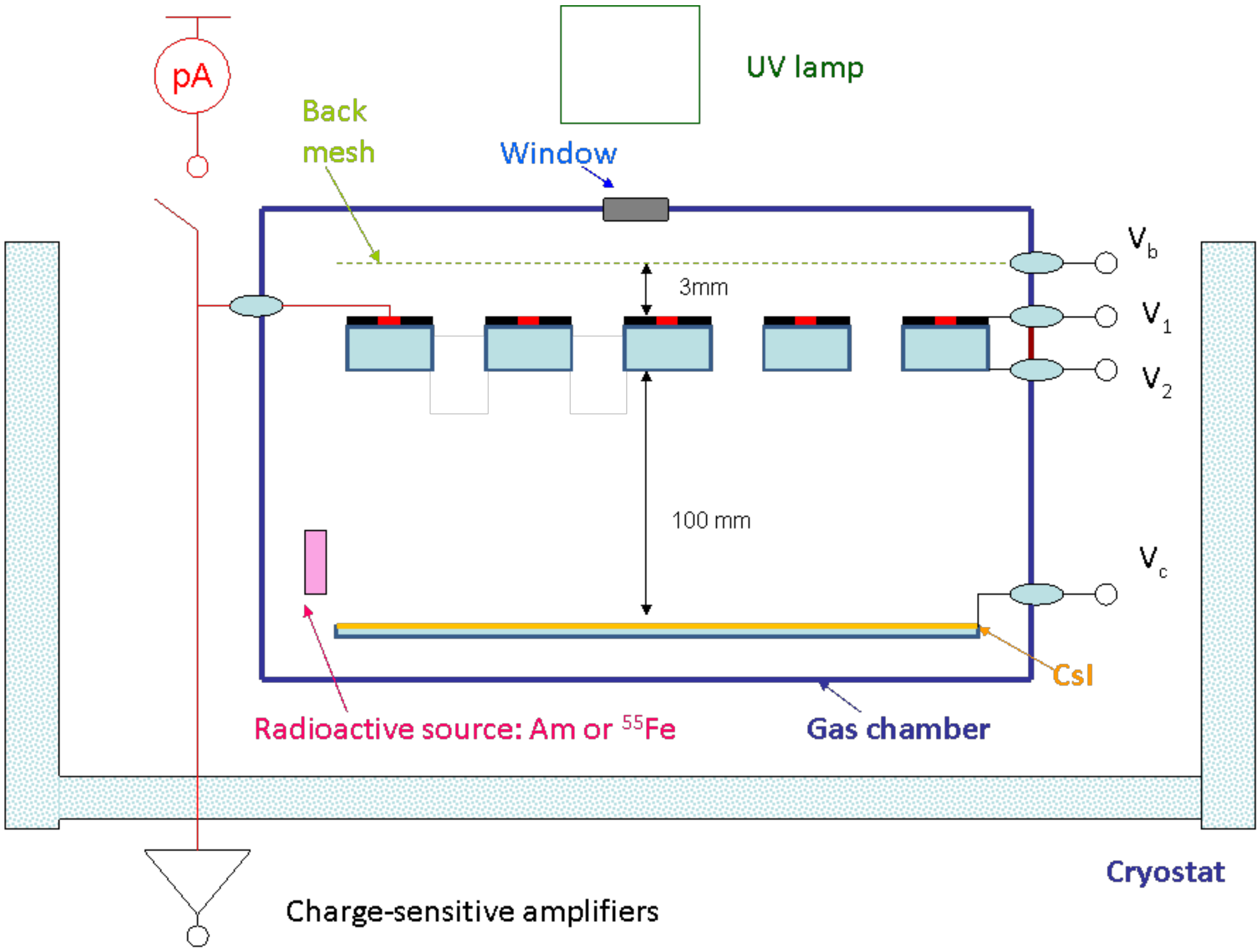

Figure 9. Schematic drawing of a simplified setup for tests and studies of RMMD combined with a CsI photocathode.

### 2.3.3. Results

Figure 10 shows the RMMD gain as a function of the overall voltage $V_{ov}$ applied to the electrodes ($V_{ov} = V_{bc}+V_{ca}$, where $V_{bc}$ is the voltage applied between the back-plane and the cathode strips -the voltage across the holes- and $V_{ca}$ is the voltage applied between the cathode strips and the anode dots), measured in Ar at room and cryogenic temperatures. The filled symbols represent the measurements performed with alpha particles, whereas the open symbols refer to the results obtained with $^{55}$Fe. Since with alpha particles one can observe signals even at a gas gain of one, the $^{241}$Am source was very convenient for precise gain measurements at gas gains $\leq 100$. At higher gains $^{55}$Fe was used and the gain values were estimated from the measured avalanche charge and the known sensitivity of the amplifier. As it can be seen, at room temperature, the maximum achievable gain was ~$3x10^4$ which is 3-10 times higher than that achieved with other micropattern gaseous detectors operated in Ar. No feedback pulses originating from the CsI photocathode were observed (they should have a delay of a few μs) even at the maximum achievable gains, proving that in RMMDs the avalanche region is geometrically well shielded with respect to the photocathode. Note that at some critical total charge in the avalanche close to $Q_{crit}$, self quenched streamers appeared (see [21] and references therein). This was shown by a jump of one order of magnitude in gas gain, observed close to some critical gains. Indeed, visual observations revealed that streamers were formed near the anode dots and propagated towards the back and drift plane. A simulation was developed describing the formation and propagation of streamers in the case of the adopted geometry of the RMMD (figures 11, 12).
Since streamers appear at some critical total charge in the avalanche, it will be interesting to check whether they are formed and at what temperature, and what the maximum achievable gain will be whenever the primary avalanches are created by single electrons. For this aim we performed measurements in current mode, when the CsI photocathode was irradiated with a Hg lamp and the photocurrent as a function of the voltage $V_{ov}$ was measured from the anode dots. Some results are depicted in figure 13. As can be seen from the curve presented in this figure, in the case of single primary photoelectrons, streamers were not observed, whereas the maximum achievable gain was, depending on the gas and temperature, between $3x10^4$ and $10^5$. These gains are high enough to detect several primary electrons that satisfy the requirements for noble liquid TPCs for dark matter searches. So far, these are the highest gains achieved at cryogenic temperatures with a single-stage gas detector.
We have also performed stability tests of the RMMD using a $^{55}$Fe source. Some results are shown in figure 14. As can be seen both at room and cryogenic temperatures the gas gain increases during the first 5-60 min (depending on the temperature), but then remains rather stable.

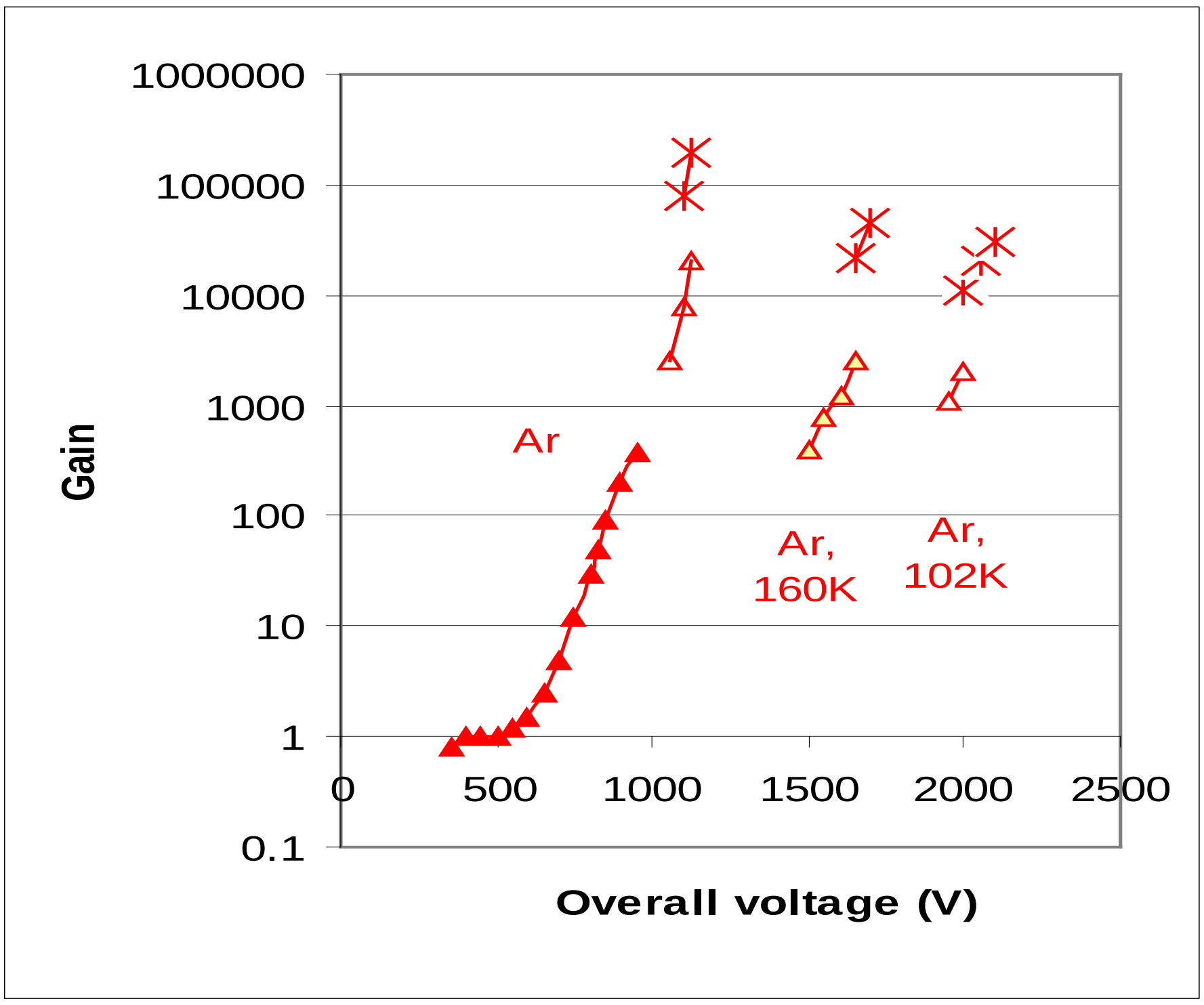

Figure 10. Gas gain curves measured in Ar at room and cryogenic temperatures. Filled symbols–alpha particles, open symbols - $^{55}Fe$. Crosses show the gas gains in self-quenched streamer mode at room temperature and at 160 and 102K.

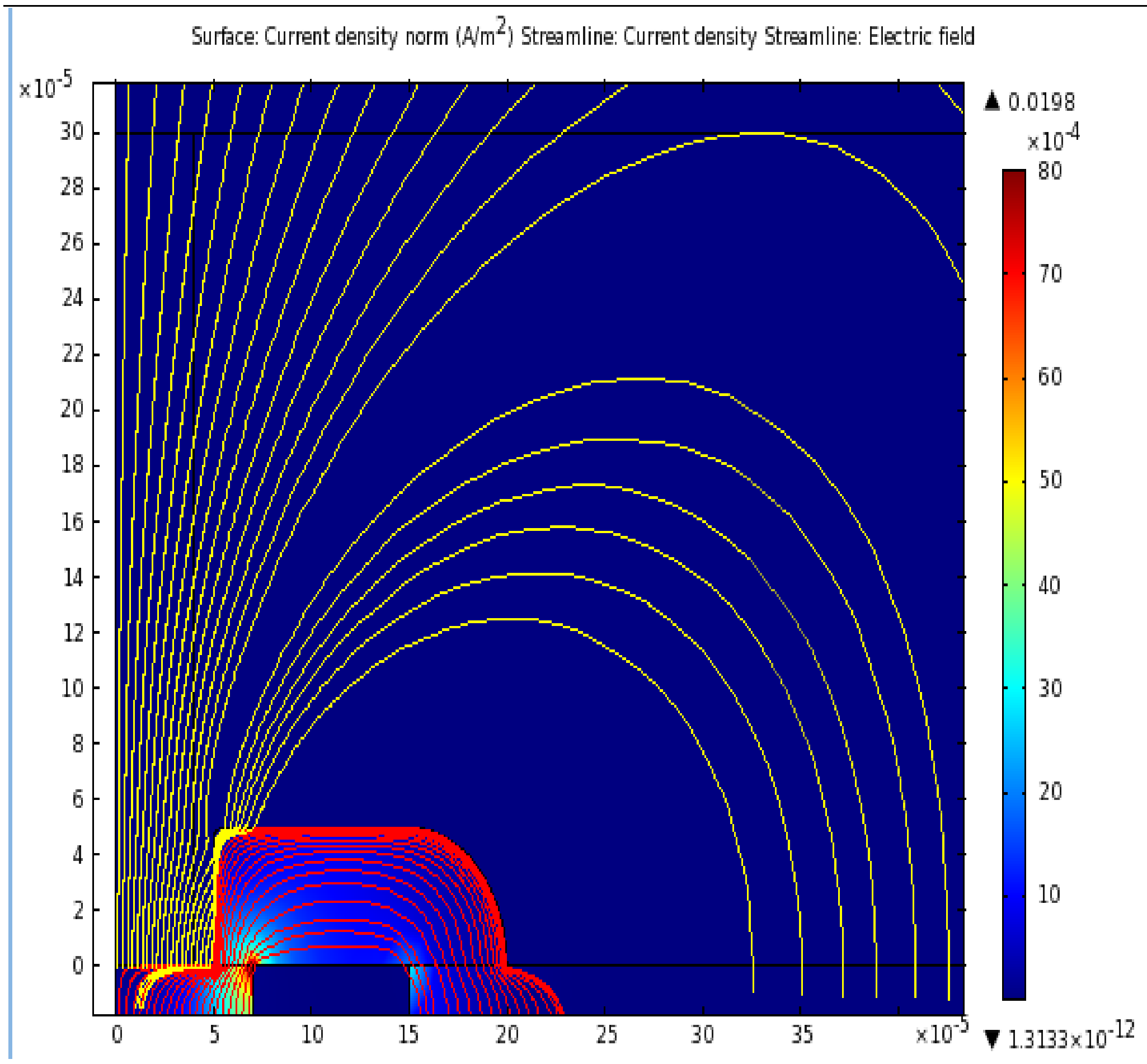

Figure 11. Simulated current lines ,assuming that the current on the Coverlay is much larger than the current due to the radiation (red lines) and electric field lines in the gas (yellow lines).

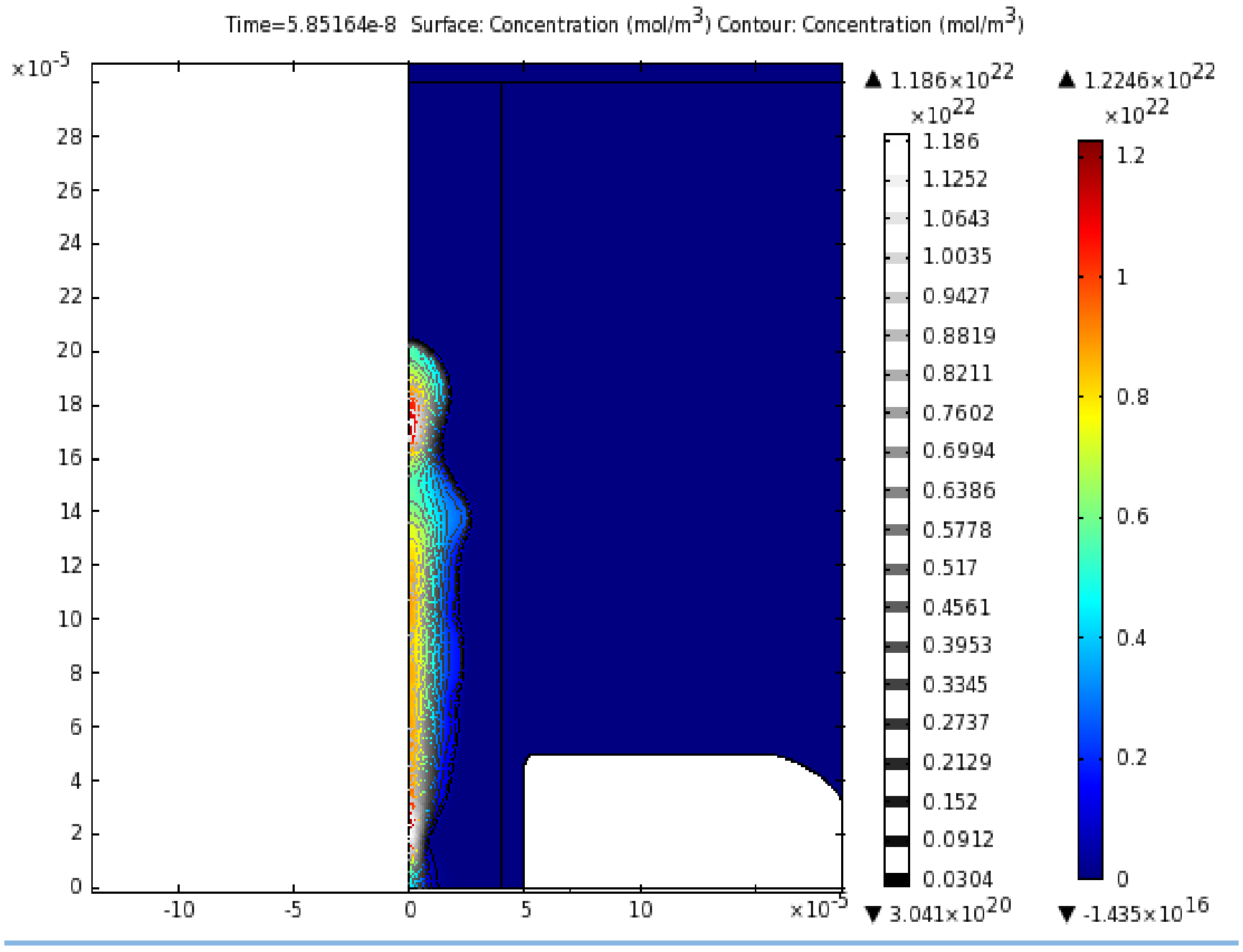

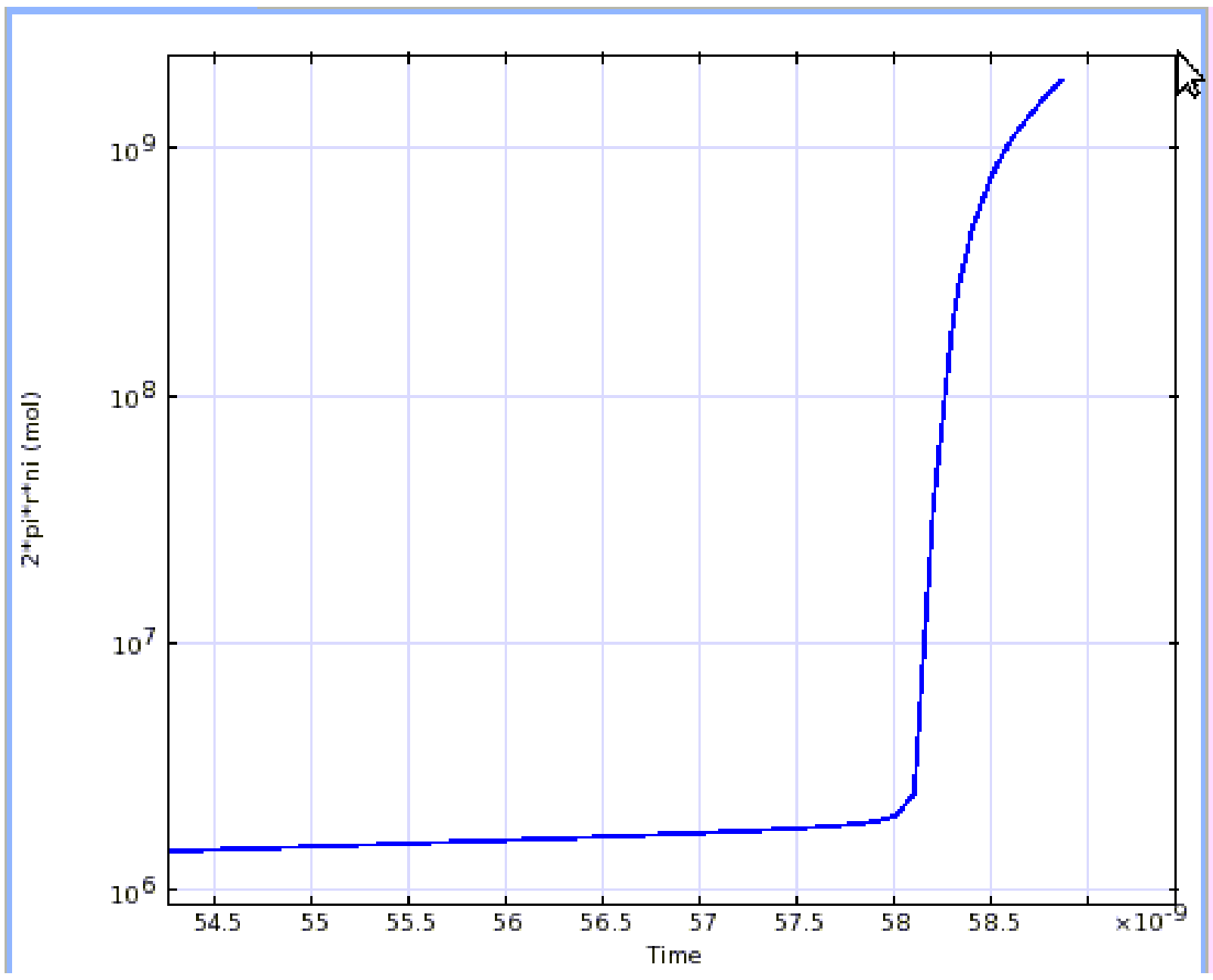

Figure 12. First results of streamer simulation (upper figure). Contours: electrons numerical density. Color wash: ion numerical density. At the beginning the current growth but then it reaches the saturation which can be an indication that the streamer becomes self-quenched (lower figure).

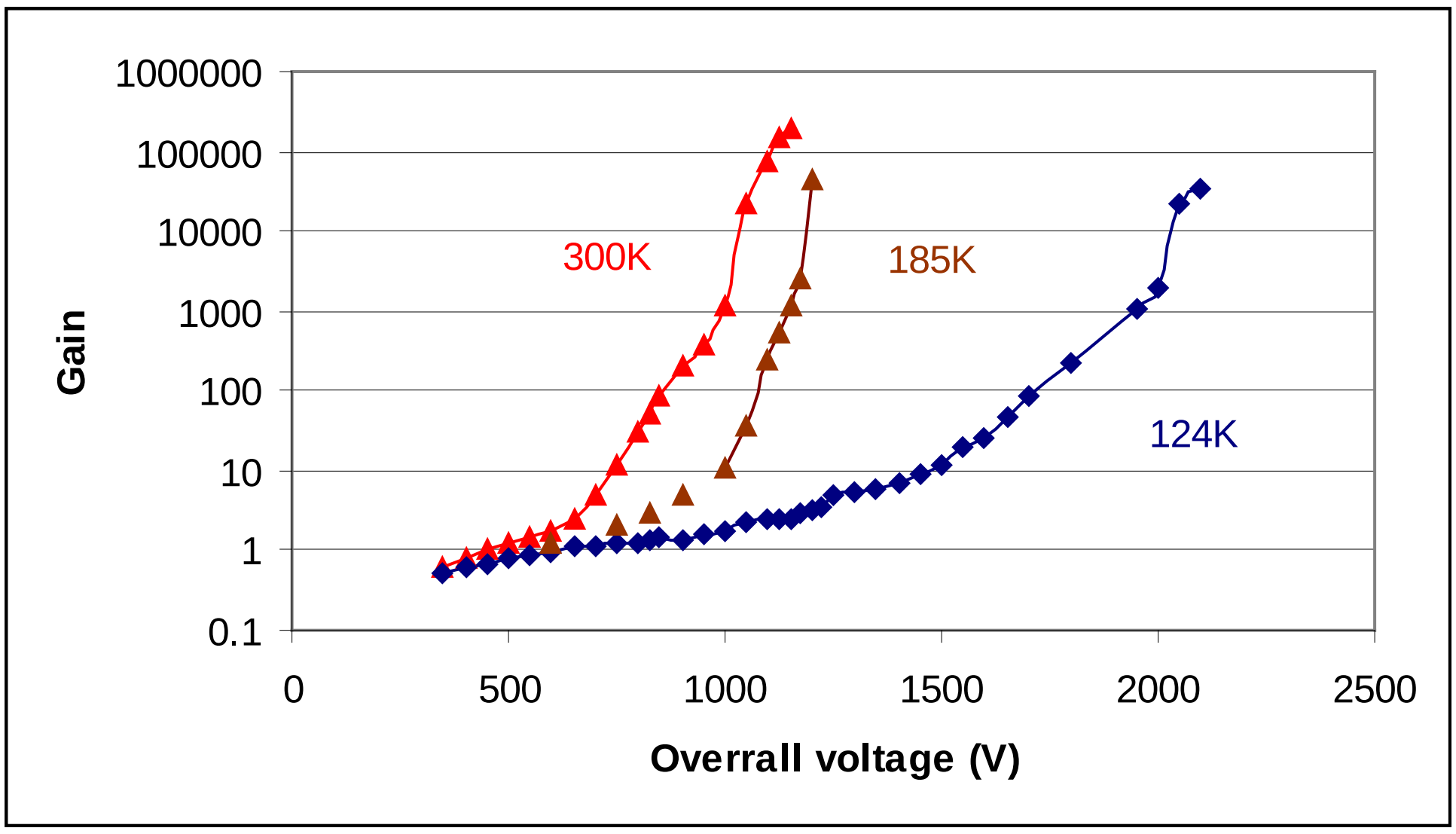

Figure 13. Gain curves of RMMD measured in current mode at various temperatures in Ar gas. The measurements at high gains were stopped as soon as first signs of current instability appeared.

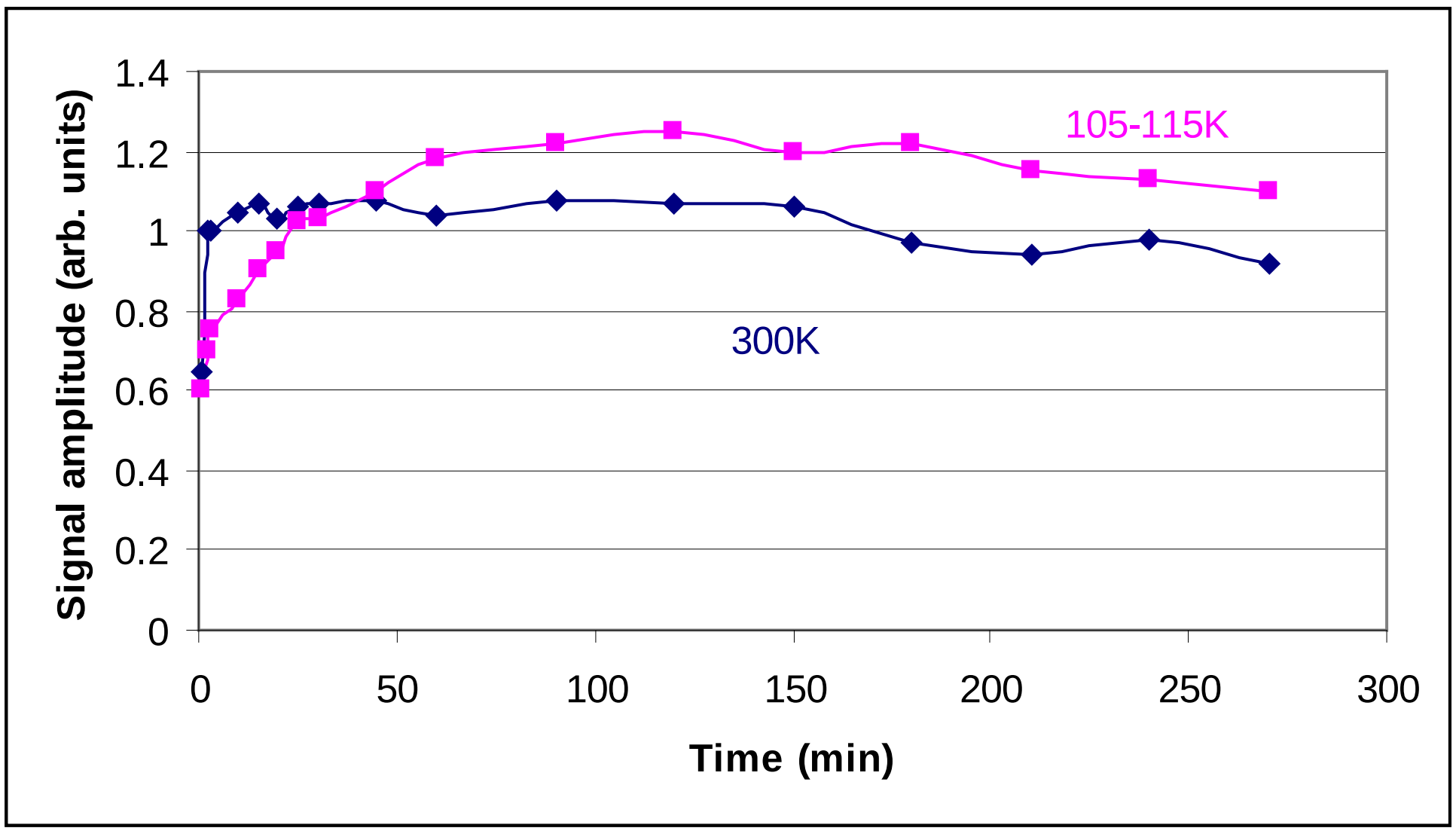

Figure 14. Results of stability tests performed with $^{55}Fe$ source at 300K and ~110K.

## 3. Microgap-Microstrips-RPC for medical imaging and other applications

### 3.1 Early developments

As was shown in [22, 23], microgap RPCs with inner fine pitch metallic strips allows a very high position resolution to be achieved: about 50 μm in digital mode. These detectors were successfully used in innovative experimental mammographic scanners. Nowadays micropattern detector technology offers new possibilities in the manufacturing of various designs of microgap detectors. In particular, one can manufacture readout strips just under the resistive layers thus increasing the robustness and the spark resistance of the detector [8]. In this work we built and tested such microgap parallel plate detectors (MMRPC) with inner strips

### 3.2. MMRPC manufacturing

The manufacturing procedure of the MMRPC is shown in figure 15. The first four steps a) to d) were very similar to the case of the RMMD described above: resistive strips with a pitch of 0.5 or 0.25 μm (depending on the specific design) and having a thickness of 35 μm were manufactured by a combination of photolithography and screen printing techniques. The difference was only in the geometrical arrangement of the strips: each resistive strip had a pick up strip positioned below it and both strips had the same width. The grooves between the resistive strips were covered with a dielectric layer to make a smooth surface (step e). Using these plates (figure 16), MMRPCs with various gap widths, *G*, were assembled: 0.5 mm, 0.18 mm and 0.118 mm (figure 17). The edges of each plate were rounded to avoid sparks in this region. As spacers, either washers, placed in the detector corners, or pillars manufactured by a photolithographic technology similarly to the case of MICROMEGAS (figure 18) were used. Note that the electrodes of this detector are somewhat similar to the anode plate of the R-MICROMEGAS, the main difference being that the cathode mesh is replaced by the cathode plate with resistive strips. This feature not only simplifies the manufacturing procedure, allowing it to be automatized, but also offers the possibility to achieve, with proper gas mixtures, high position and high time resolutions at the same time, similar to some designs of timing RPCs.

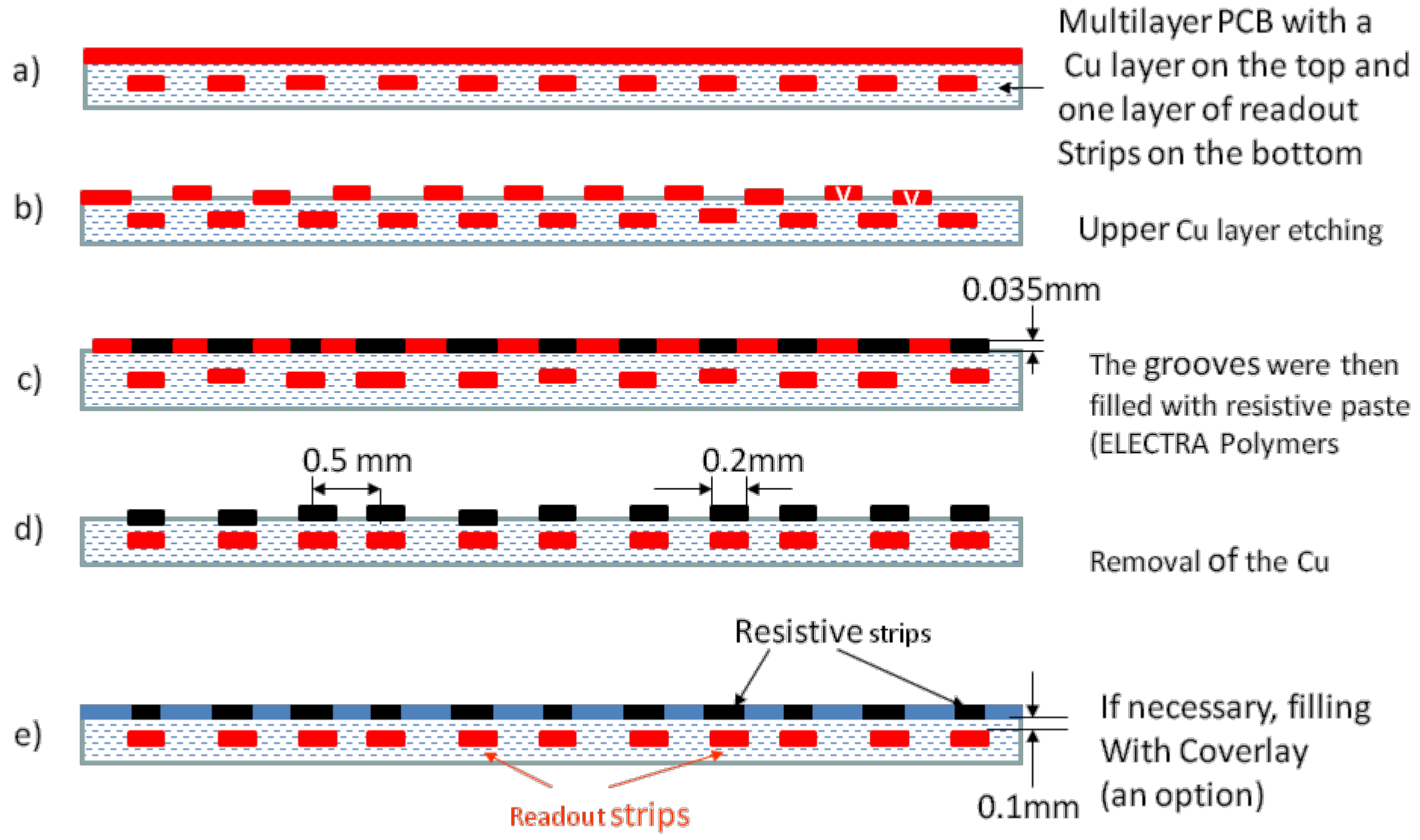

Fig.15. Manufacturing steps of MMRPC.

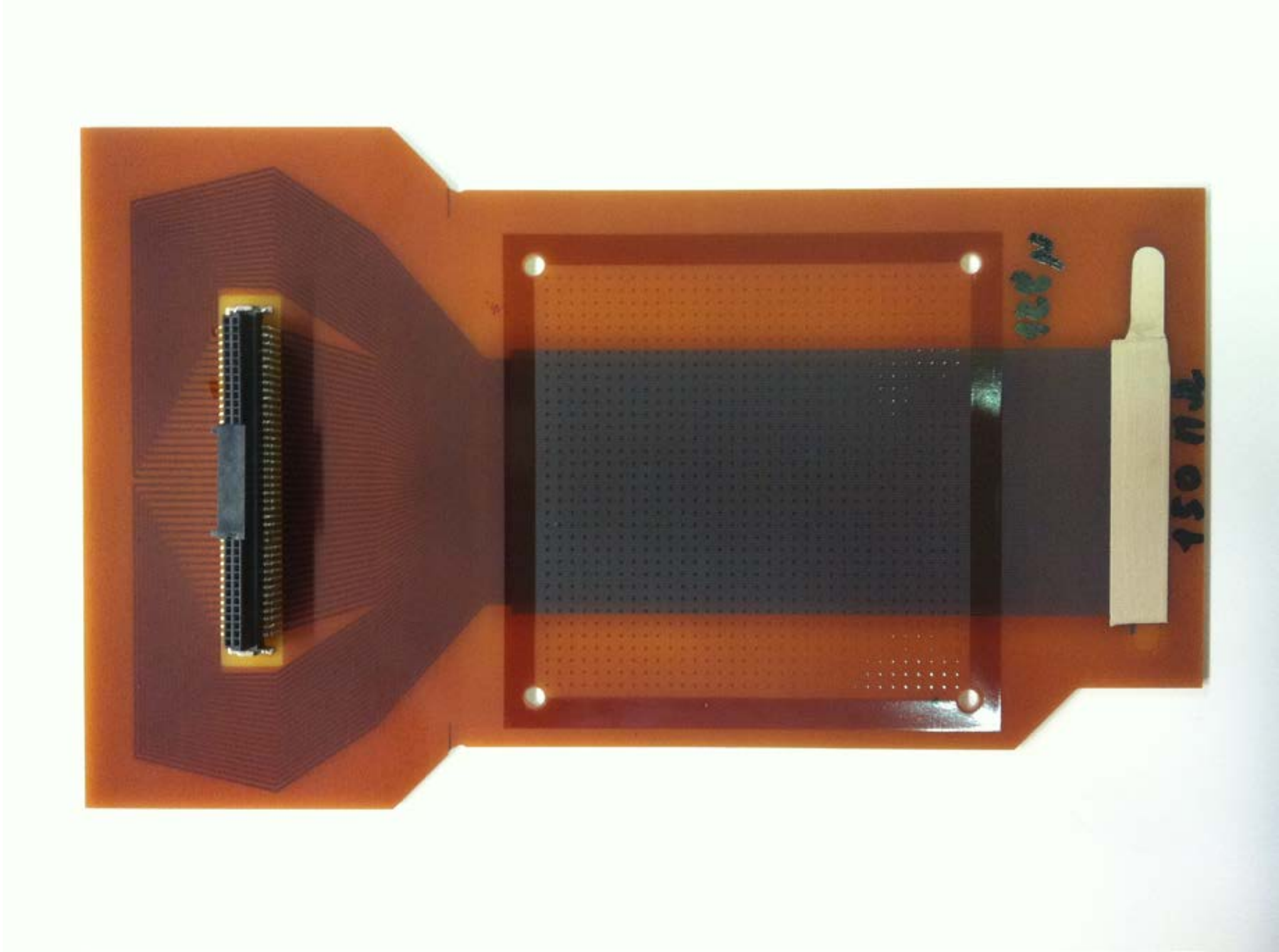

Figure 16. Photograph on one of the resistive plates used for the assembling of the MMRPC. Dots seen in the active part of the detector are spacers manufactured by photolitographic technology developed in the past for the production of resistive MICROMEGAS.

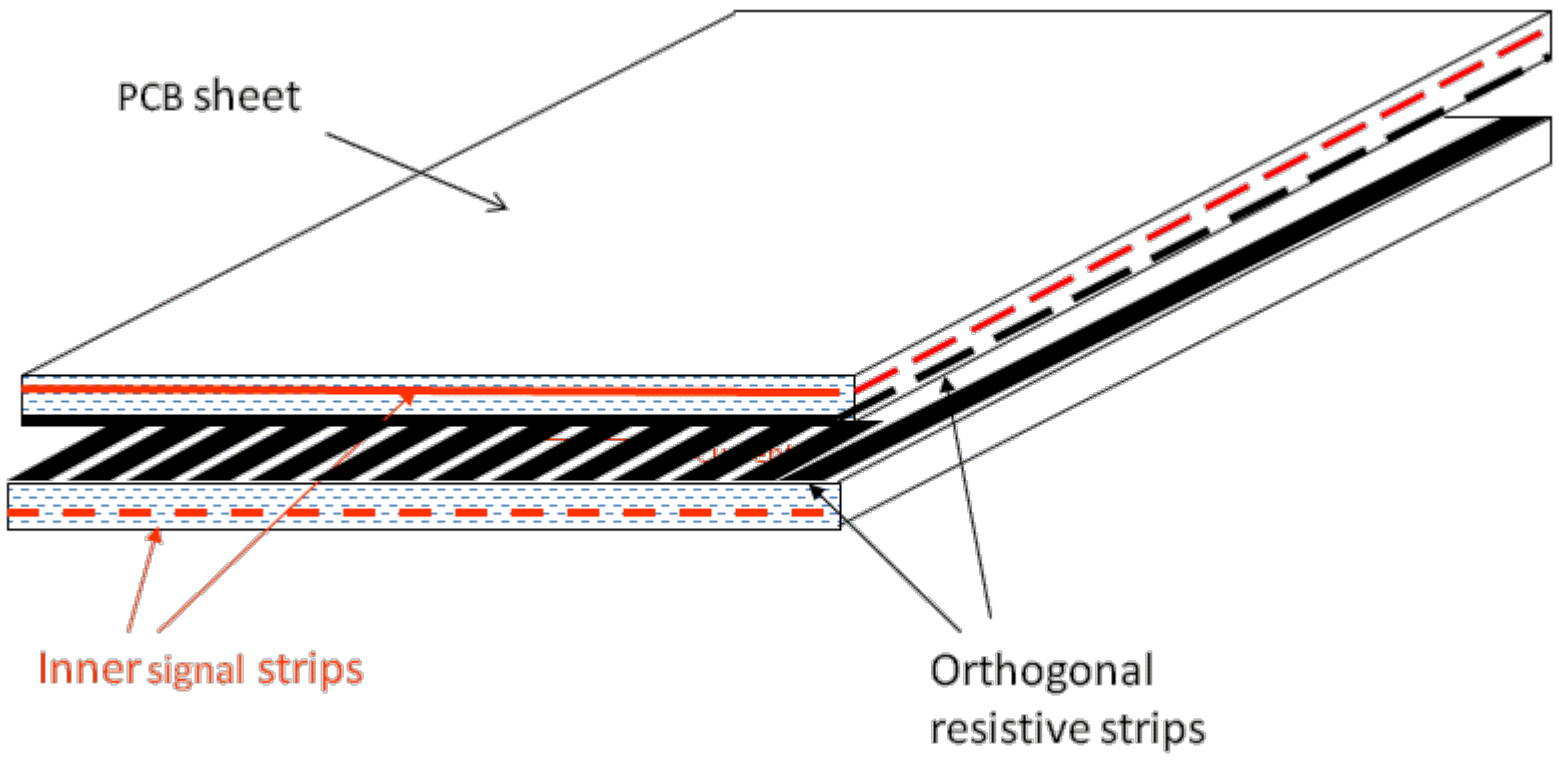

Figure 17. Artistic view of the assembled MMRPC.

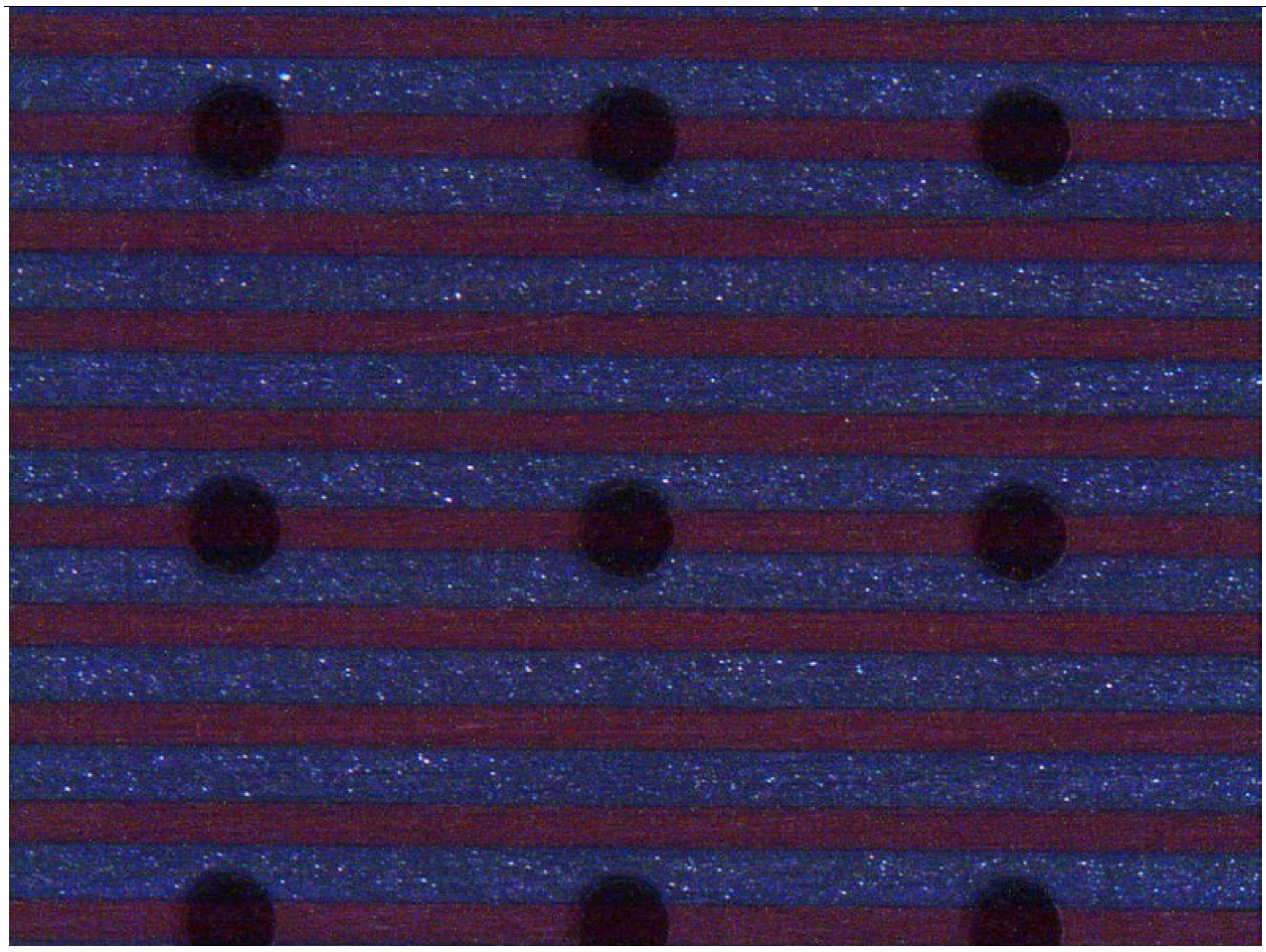

Figure 18. Magnified photograph of the inner surface of the MMRPC electrode. One can see the resistive strips and spacers (black dots).

## 3.3 Experimental setup for testing MMRPC

The detectors were tested using the experimental set up schematically shown in figure 19. The primary electrons in the detector volume were created using an X-ray gun (depending on measuremenst either Eclips, Oxford Instr. or XTF5011 Oxford Intr. were used) or by an Hg UV lamp. To accurately evaluate the position resolution of the detector, slit collimators of width 0.1 mm or 0.05 mm oriented parallel to the MMRPC electrode and close to its cathode were used (figure 20). For the collimation in the direction perpendicular to the MMRPC electrodes a vertical collimator was used of width 0.05 mm. In some measurements the cathode of the MMRPC was coated with a CsI layer (figure 20). When the gap between the electrodes was illuminated with the UV light a photocurrent was created allowing gain measurements in current mode to be performed.

Tests were done in Ne, Ar and their mixtures with $CO_2$ at a total pressure of 1atm.

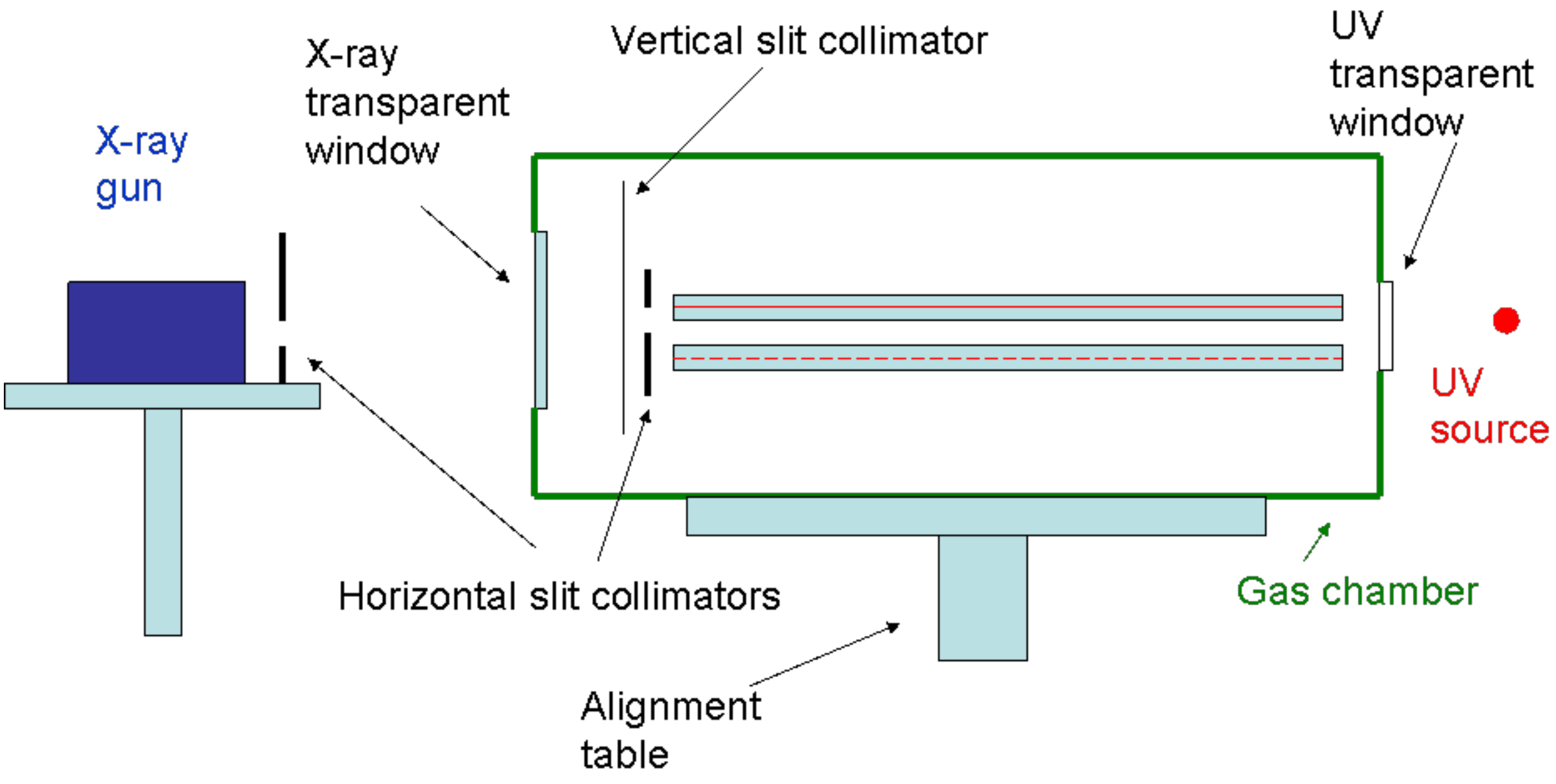

Figure 19. Schematic drawing of the experimental setup for tests of the MMRPC.

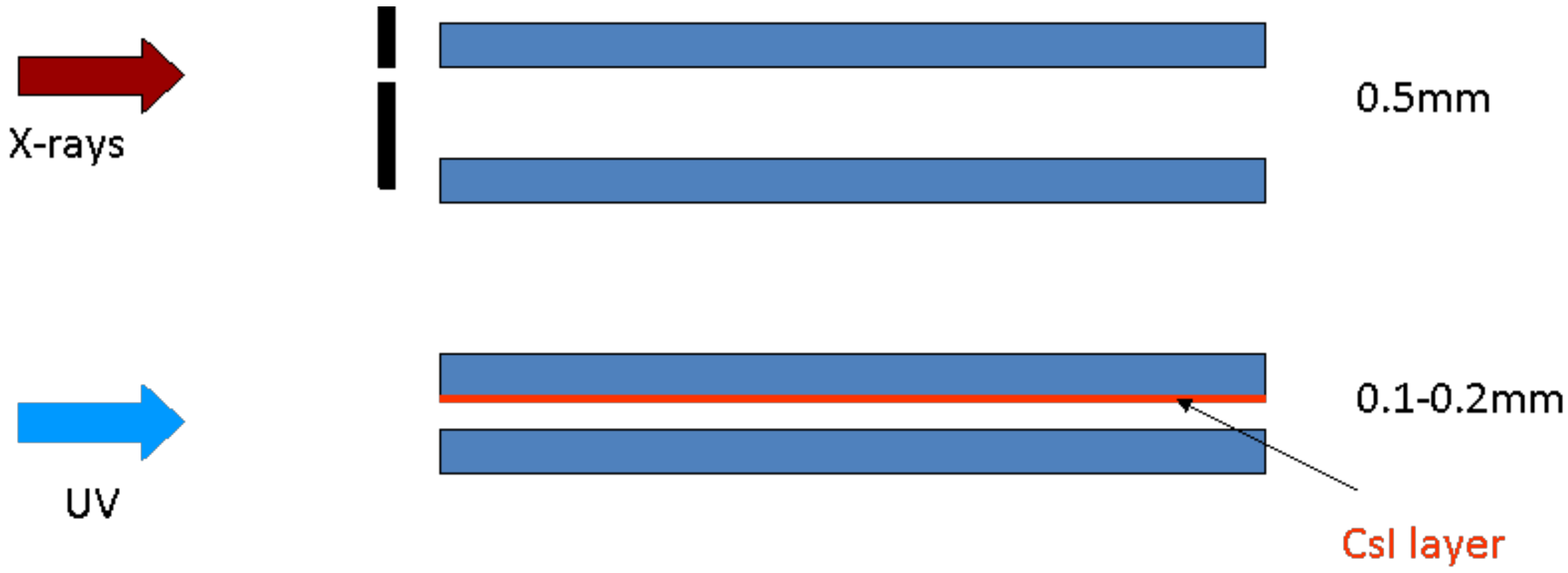

Figure 20. Schematic drawing explaining measurements with collimated X-rays and with UV light.

### 3.4. Results

Figure 21 shows the gas gain of the MMRPC measured both in current and pulse mode, in Ar+$CO_2$ gas mixtures. These mixtures were chosen in order to compare the gas gain of the MMRPC with the gas gains of other micropattern detectors, for example with R-MICROMEGAS (see [8]). Note that in contrast to RMMDs, which have a drift region, it is not straightforward to measure the gas gain in MMRPCs, where no drift region is present. One way to measure the gas gain of the MMRPC is to let the radiation enter the detector volume at a well known distance from the anode and then apply the equation of gas gain in parallel-plate geometry, $A=exp(\alpha d)$, where $\alpha$ is the Townsend coefficient and $d$ is the distance of the primary electron from the anode. The best way to measure *A* is to use the surface photoelectric effect from the cathode (in this case *d*=*G*-the cathode–anode distance) and perform measurements in current mode at low gains and in pulse mode at high gains. Alternatively it is possible to use a well-collimated X-ray beam entering the detector volume close to the cathode and parallel to it. However, for technical reasons this can be used only when the gap is large enough, equal to or greater than 0.5 mm. In this work, we exploited both options. In the case of the measurements based on the photoelectric effect, we coated the cathode plate of the MMRPC with a thin CsI layer by a spray technique [24]. The UV beam of the Hg lamp was shot inside the MMRPC almost parallel to its electrodes.

In Figure 21 gain curves measured in current mode are marked with solid symbols and in pulse mode with open symbols. In the case of G=0.5 mm, we performed measurements with UV and X-rays whereas at smaller gaps (G=0.18 and 0.118 mm) with UV only. As can be seen, gas gains above $10^6$ were achieved with all gaps. This is almost 100 times higher than with R-MICOMEGAS.

Fig. 22 depicts results of the measurements of the avalanche-induced charge profiles on the strips. As can be seen, the FWHM of this distribution is about 0.5 mm. Note that earlier, with a different design of a narrow gap strip RPC orientated towards medical applications, we have already achieved a position resolution of about 50 μm in digital mode. More accurately, the position resolution of the MMRPC will be measured during the oncoming test with charged particle beams when we plan to use standard RPC gas mixtures (90% $C_2H_2F_4$ + 10%$SF_6$ or 90% $C_2H_2F_4$ + 5% $SF_6$ + 5%$C_4H_{10}$) in order to achieve at the same time a high position resolution and high time resolution, typical of small gap RPCs.

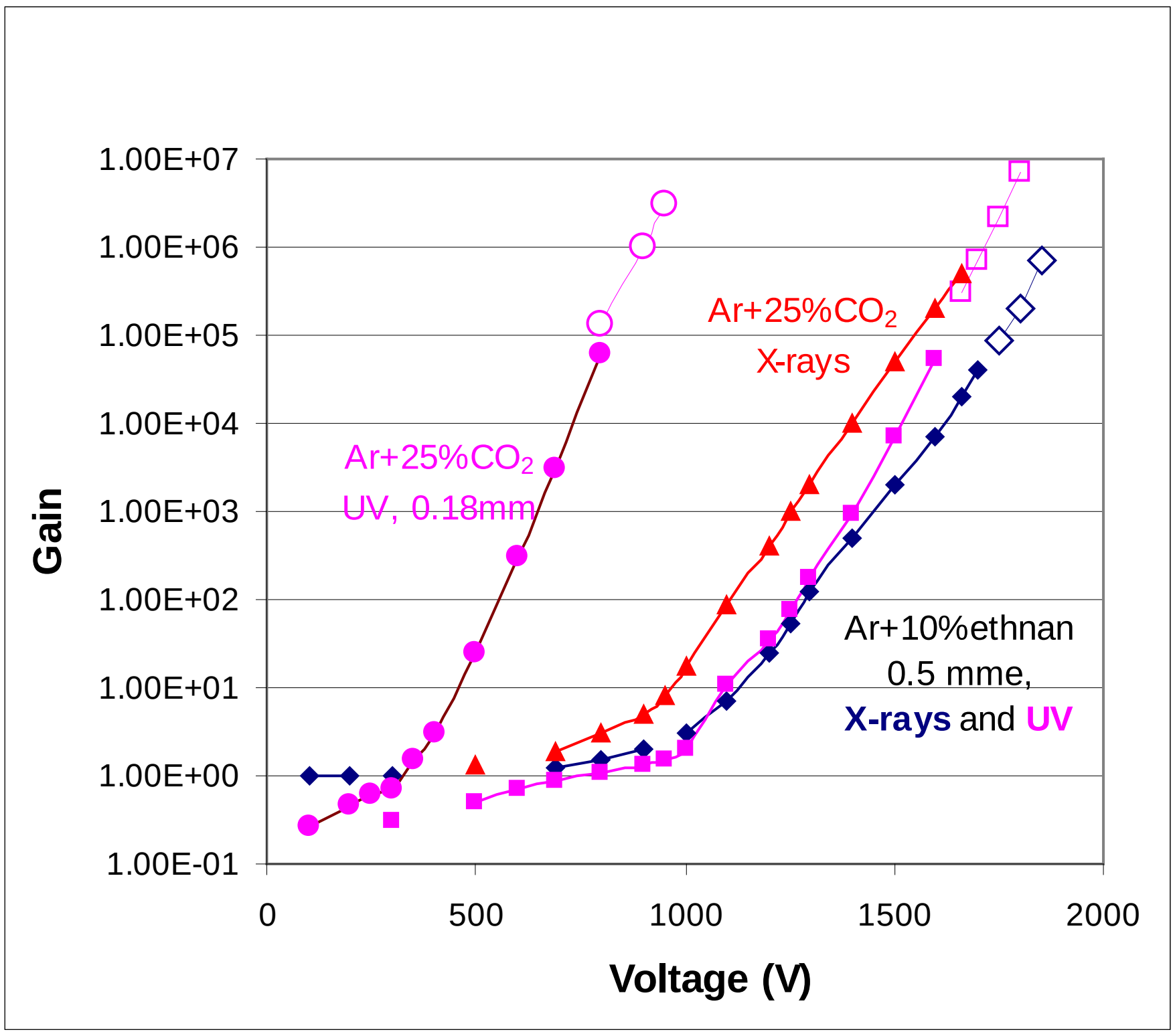

Figure 21. Gain vs. voltage curves as measured with a MMRPC, having gaps G=0.18 mm (circles) and G=0.5 mm (triangles, squares and rhombuses) in current mode (filled symbols) and in pulse mode (open symbols); circles and squares-measurements with UV.

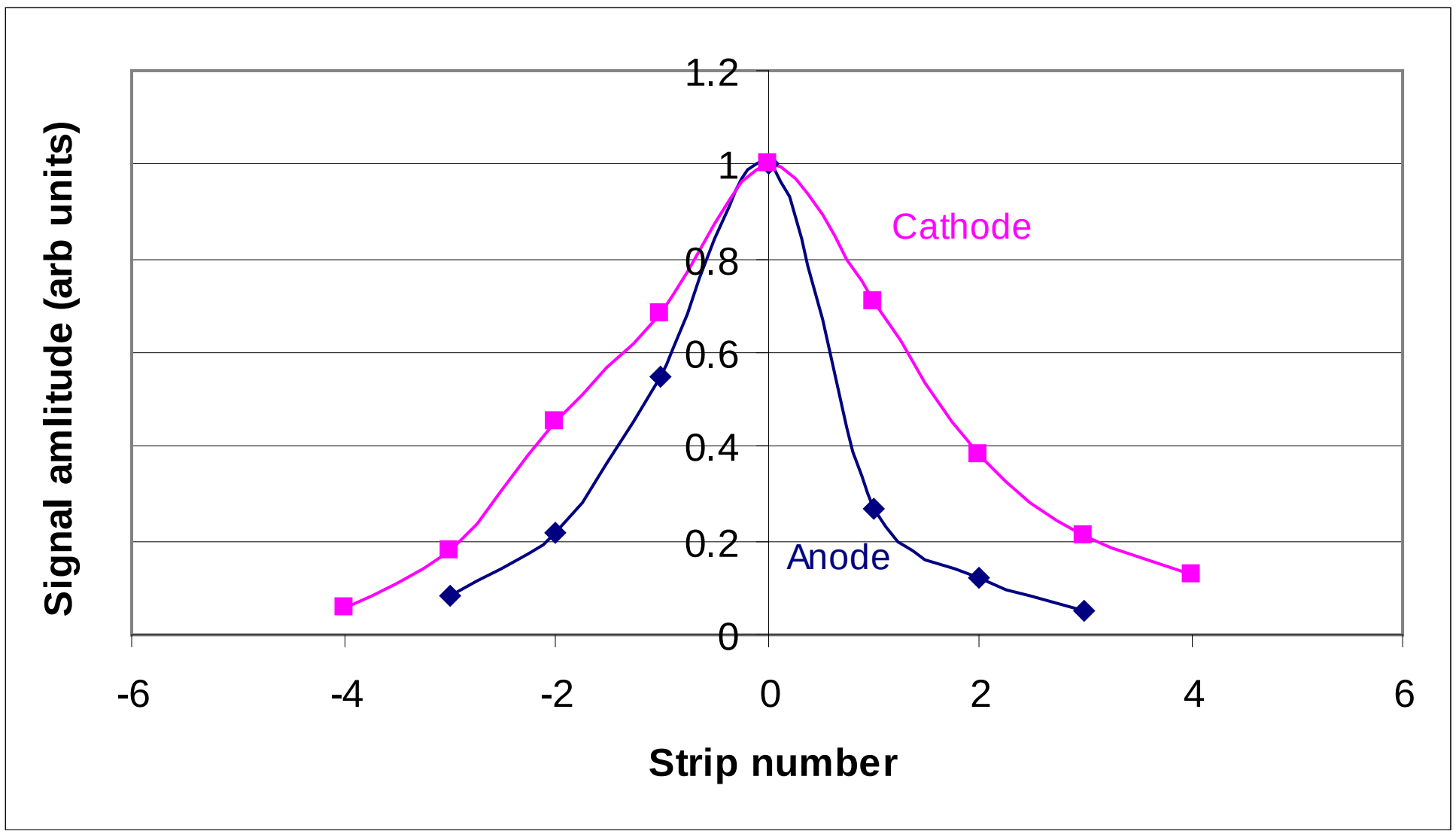

Figure 22. Induced charge profile measured with collimated X-ray beam entering the MMRPC close to its cathode and parallel to it. The horizontal and vertical collimator widths were 0.1mm and strip pitches 250 μm.

## 4. Discussion and Outlook

Two new designs of micropattern gaseous detectors with resistive electrodes have been studied: the microdot-microhole detector and the microgap-microstrip-RPC. These novel detectors have several important advantages over the conventional micropattern detectors with metallic electrodes like GEM and MICROMEGAS. For example, the proposed detectors feature higher values of the maximum achievable gas gains, a simpler layout with fewer components, a simpler production technique and cost effectiveness.  Our studies have revealed many interesting features of such detectors, for example, the microdot detector can operate in self quenched streamer mode and at higher voltages in streamer mode similarly to RPCs.

The technology of manufacturing microgap RPC with inner readout strips has been developed. The FWHM of the profile of the induced signal on the readout strips measured with 0.1 mm wide collimators is around 0.5 mm indicating that the intrinsic position resolution of the MMRPC is even better. The main application of this detector is low dose x-ray scanners.

It is known that small gap RPCs in some gas mixtures, for example 90% $C_2H_2F_4$ + 10%$SF_6$ or 90% $C_2H_4F_2$ + 5% $SF_6$ + 5%$C_4H_{10}$, exhibit high time resolution. We thus believe that our microgap RPCs may combine high position and time resolutions and thus can be used for such applications as TOF- PET or tracking of charged particles [9, 23,25-27] etc.

The general conclusion is that resistive micropattern detectors, described in this and earlier publications [5] have a great potential and thus are very promising for many applications.